\documentstyle[11pt,aaspp]{article}
\input epsf
\def\l*{$L_*$\/}
\def\etal{{\it et al. }}

\def\kms{\rm ~km~s^{-1}}

\def\kmsmpc{km s$^{-1}$ Mpc$^{-1}$\ }

\def\l*{$L_{*}$}

\def\gsim{ \lower .75ex \hbox{$\sim$} \llap{\raise .27ex \hbox{$>$}} }
\def\lsim{ \lower .75ex \hbox{$\sim$} \llap{\raise .27ex \hbox{$<$}} }
\def\pp{\noindent\parshape 2 0truecm 16.0truecm 2.0truecm 15truecm}

\def\spose#1{\hbox to 0pt{#1\hss}}
\def\simlt{\mathrel{\spose{\lower 3pt\hbox{$\mathchar"218$}}
     \raise 2.0pt\hbox{$\mathchar"13C$}}}
\def\simgt{\mathrel{\spose{\lower 3pt\hbox{$\mathchar"218$}}
'     \raise 2.0pt\hbox{$\mathchar"13E$}}}

\font\titlefont=cmss17

\begin{document}



\title{\titlefont On the Survival and Destruction of Spiral Galaxies in Clusters}


%

\vskip 0.5truecm

\centerline{\bf Ben Moore$^{1}$, George Lake$^{2}$, Thomas Quinn$^{2}$ \& Joachim Stadel$^{2}$}

\

\centerline{{\bf 1.} Department of Physics, University of Durham, Durham City, DH1 3LE, UK}

\centerline{{\bf 2.} Department of Astronomy, University of Washington, Seattle, WA 98195, USA}

\


 
\begin{abstract}

We follow the evolution of disk galaxies within a cluster that forms
hierarchically in a cold dark matter N-body simulation.  At a
redshift $z=0.5$ we select several dark matter halos that have quiet
merger histories and are about to enter the newly forming cluster
environment.  The halos are replaced with equilibrium high resolution
model spirals that are constructed to represent examples of low
surface brightness (LSB) and high surface brightness (HSB) galaxies.
Varying the disk and halo structural parameters reveals that the
response of a spiral galaxy to tidal encounters depends primarily on
the potential depth of the mass distribution and the disk scale
length.  LSB galaxies, characterised by slowly rising rotation curves
and large scale lengths, evolve dramatically under the influence of
rapid encounters with substructure and strong tidal shocks from the
global cluster potential \--- galaxy harassment. We find that up to
90\% of their stars are tidally stripped and congregate in large
diffuse tails that trace the orbital path of the galaxy and form the
diffuse intra-cluster light.  The bound stellar remnants closely
resemble the dwarf spheroidals (dE's) that populate nearby clusters.
HSB galaxies are stable to the chaos of cluster formation and tidal
encounters. These disks lie well within the tidally limited dark
matter halos and their potentials are more concentrated.
Although very few stars are stripped, the scale height of the disks
increases substantially and no spiral features remain, therefore
we speculate that these galaxies would be identified as S0
galaxies in present day clusters.



\end{abstract}


\vfil\eject

\section{Introduction}

Clusters of galaxies provide a unique environment wherein the galaxy
population has been observed to rapidly evolve over the past few
billion years (Butcher \& Oemler 1978, 1984, Dressler \etal 1998).  At
a redshift $z \gsim 0.4$, clusters are dominated by spiral galaxies
that are predominantly faint irregular or Sc-Sd types. Some of these
spirals have disturbed morphologies; many have high rates of
star-formation (Dressler \etal 1994a).  Conversely, nearby clusters
are almost completely dominated by dwarf spheroidal (dSph,dE),
lenticulars (S0) and elliptical galaxies (Bingelli \etal 1987, 1988,
Thompson \& Gregory 1993).  Observations suggest that the elliptical
galaxy population was already in place at much higher redshifts, at
which time the S0 population in clusters is deficient compared to
nearby clusters (Couch \etal 1998, Dressler \etal 1998).  This
evolution of the morphology-density relation appears to be driven by
an increase in the S0 fraction with time and a corresponding decrease
in the luminous spiral population.

Low surface brightness (LSB) galaxies appear to avoid regions of high galaxy
densities (Bothun \etal 1993, Mo \etal 1994).  This is somewhat puzzling since
recent work by Mihos \etal (1997) showed that LSB disk galaxies
are actually {\it more} stable to tidal encounters than
HSB disk galaxies.  In fact, LSB galaxies have lower disk mass surface
densities and higher mass-to-light ratios, therefore their disks are
less susceptible to internal global instabilities, such as bar
formation.  However, in a galaxy cluster, encounters occur frequently and 
very rapidly, on a shorter timescale than investigated by Mihos \etal and
the magnitude of the tidal shocks are potentially very large. 

Several physical mechanisms have been proposed that can strongly
affect the morphological evolution of disks: ram-pressure stripping
(Gunn \& Gott 1978), galaxy merging (Icke 1985, Lavery \& Henry 1988, 1994) and
galaxy harassment (Moore \etal 1996a, 1998). The importance of these
mechanisms varies with environment: mergers are frequent in groups but
rare in clusters (Ghigna \etal 1998), ram pressure removal of gas is
inevitable in rich clusters but will not alter disk morphology (Abadi
\etal 1999).  The morphological transformation in the dwarf galaxy
populations ($M_b>-16$) in clusters since $z=0.4$ can be explained by
rapid gravitational encounters between galaxies and accreting
substructure \-- galaxy harassment.  The impulsive and resonant
heating from rapid fly-by interactions causes a transformation from
disks to spheroidals.  

The numerical simulations of Moore \etal focussed on the evolution of
fainter Sc-Sd spirals in static clumpy cluster-like potentials and
their transition into dSph's.  In this paper we examine the role of
gravitational interactions in driving the evolution of luminous
spirals in dense environments.  We shall use more realistic
simulations that follow the formation and growth of a large cluster
that is selected from a cosmological simulation of a critical CDM
universe.  The parameter space for the cluster model is fairly well
constrained once we have adopted hierarchical structure formation.
The structure and substructure of virialised clusters is nearly
independent of the shape and normalisation of the power
spectrum. Clusters that collapse in low Omega universes form earlier,
thus their galaxies have undergone more interactions. The cluster that
we follow virialises at $z\sim 0.3$, leaving about 4 Gyrs for the
cluster galaxies to dynamically evolve.

The parameter space for the model spirals is much larger. Mihos \etal
examined the effects of a single encounter at a fixed number of disk
scale lengths, whilst varying the disk surface brightness and keeping
other properties fixed. The key parameter that determines whether or
not dark matter halos survive within a cluster N-body simulation is
the core radius of the substructure, which is typically dictated by
the softening length (Moore \etal 1996b).  We suspect that
this may also be a key factor that governs whether or not a given disk
galaxy will survive within a dense environment. Most LSB galaxies have
slowly rising rotation curves indicating ``soft'' central potentials
and thus should be more unstable than HSB galaxies with flat rotation
curves.  To investigate this hypothesis we constructed several
different galaxy models. A ``typical'' luminous HSB and LSB disk
galaxy that both lie on the same point in the Tully-Fisher relation,
as well as a sequence of models that have different surface
brightness, disk scale lengths and halo structural parameters.

In Section 2 we examine the response of three different model disk
galaxies to a single strong tidal encounter.  Section 3 discusses the
cosmological simulations in which we follow the hierarchical evolution
of the mass distribution.  In Section 4 we isolate the properties that
determine the stability or instability of disk galaxies orbiting
within a cluster and summarise our results in Section 5.

\section{The response to strong impulsive encounters}

For a given orbit through a cluster, the visible response of a disk
galaxy to a tidal encounter depends primarily upon its internal
dynamical timescale.  Galaxies with cuspy central mass distributions,
such as ellipticals, have short orbital timescales at their centres
and they will respond adiabatically to tidal perturbations.  Sa-Sb
spirals have flat rotation curves, therefore a tidal encounter will
cause an impulsive disturbance to a distance $\sim v_c b/V$ from its
centre, where $b$ is the impact parameter, $V$ is the encounter
velocity and $v_c$ is the galaxy's rotation speed. LSB galaxies and
Sc-Sd galaxies have slowly rising rotation curves, indicating that the
central regions are close to a uniform density.  The central dynamical
timescales are constant throughout the inner disk and an encounter
that is impulsive at the core radius will be impulsive throughout the
galaxy.

Galaxy-galaxy encounters within a virialised cluster occurs at a
relative velocity $\sim \sqrt 2\sigma_{1d}$. Substituting parameters
for an Sa--Sb galaxy, such as the Milky Way, orbiting within a cluster
we find that such encounters will not perturb the disk within $\sim
3r_d=10$ kpc.  However, tidal shocks from the mean cluster field also
provides a significant heating source for those galaxies on eccentric
orbits (Byrd \& Valtonen 1990, Valluri 1993, Moore \etal 1996a).
Ghigna \etal (1998) studied the orbits of several hundred dark halos
within a cluster that formed hierarchically in a cold dark matter
universe. The median ratio of apocenter to pericenter was 6:1, with a
distribution skewed towards radial orbits.  More than 20\% of the
halos were on orbits more radial than 10:1. A galaxy on this orbit
would move past pericenter at several thousand $\kms$ and would be
heated across the entire disk.  We shall examine the response of a
disk to a single impulse encounter using N-body simulations.

\subsection{The model spiral galaxies}

We use the technique developed by Hernquist (1989) to construct
equilibrium spiral galaxies with disk, bulge and halo components,
designed to represent ``standard'' HSB and LSB disk galaxies.  All the
model galaxies investigated in this paper have rotation curves that
peak at $200\kms$ and would therefore be a little less luminous than
``$L_*$'' spirals. Each model has an exponential disk with scale
length $r_d=3$ kpc or $r_d=7$ kpc and a scale height $r_z=0.1r_d$. The
disks are all constructed using 20,000 star particles and are are
stable with a Toomre ``Q'' parameter of 1.5.  Each galaxy has a dark
matter halo constructed using 50,000 particles distributed as a
modified isothermal sphere with core radius, $r_h=1.5$ kpc or $r_h=10$
kpc.  (Here, $r_h$ is the radius at which the contribution to
the rotation curve is 0.7 times the peak contribution.)
One of the ``HSB'' model galaxies has a small bulge with mass
25\% of the disk mass. 

To examine the effect of a single impulsive tidal encounter we
constructed three seperate model galaxies.  We have specifically
constructed two of these model galaxies so that they both lie at the
same point on the Tully-Fisher relation, yet the galaxies will have
different internal mass distributions (Zwaan \etal 1995, de Blok \&
McGaugh 1997). Figure 1 shows the contribution to the rotation
velocity of the disks from each component.  The galaxy in Figure 1(a)
has a concentrated mass distribution as indicated by the flat rotation
curve.  Note that the bulge component of the HSB galaxy has ensured
that the rotation curve is close to flat over the inner 7 disk scale
lengths, and is fairly typical of the mass distribution of HSB
galaxies (Persic \& Salucci 1997). The galaxy in Figure 1(b) has a
larger disk scale length and a rotation curve that rises slowly in the
central region, typical of that measured for LSB galaxies (de Blok \&
McGaugh 1996, de Blok \& McGaugh 1997).

Although some giant LSB galaxies have a bulge component, their
rotation curves still rise slowly. For example, 3 of the 4 LSB
galaxies observed by Pickering \etal (1997), with $v_c\gsim 200\kms$
(NGC 7589, F586-6 and Malin 1) all have rotation curves that rise more
slowly than than our standard example in Figure 1b.  However, in
Section 4 we also study the case of extended LSB disks in more
concentrated potentials. Giant LSB galaxies also have disk scale
lengths $\sim 10\--20$ kpc ({\it e.g.} Figure 9 of Zwaan \etal),
larger than the conservative value we adopt here. We shall see later
that galaxies with larger scale lengths are more unstable to
disruption and transformation to dSph.

Our model galaxies are somewhat different to those used by Mihos \etal
(1997), who kept the scale lengths constant and only varied the disk
mass surface density.  In order to examine the effects of surface
density we construct a third ``LSB'' model with a disk 1/8th of the
mass of the previous model as shown in Figure 1(b).  The HSB model and
two LSB models have disk masses of $4\times10^{10}M_\odot$,
$4\times10^{10}M_\odot$ and $5\times10^{9}M_\odot$
respectively. Adopting a B band mass to light ratio of 2, the central
surface brightness of the HSB galaxy is 20.6 mags arcsec$^{-2}$,
whilst that of the LSB models are 22.5 and 25.7 mags arcsec$^{-2}$.

\begin{figure}
\centering
\epsfxsize=\hsize\epsffile{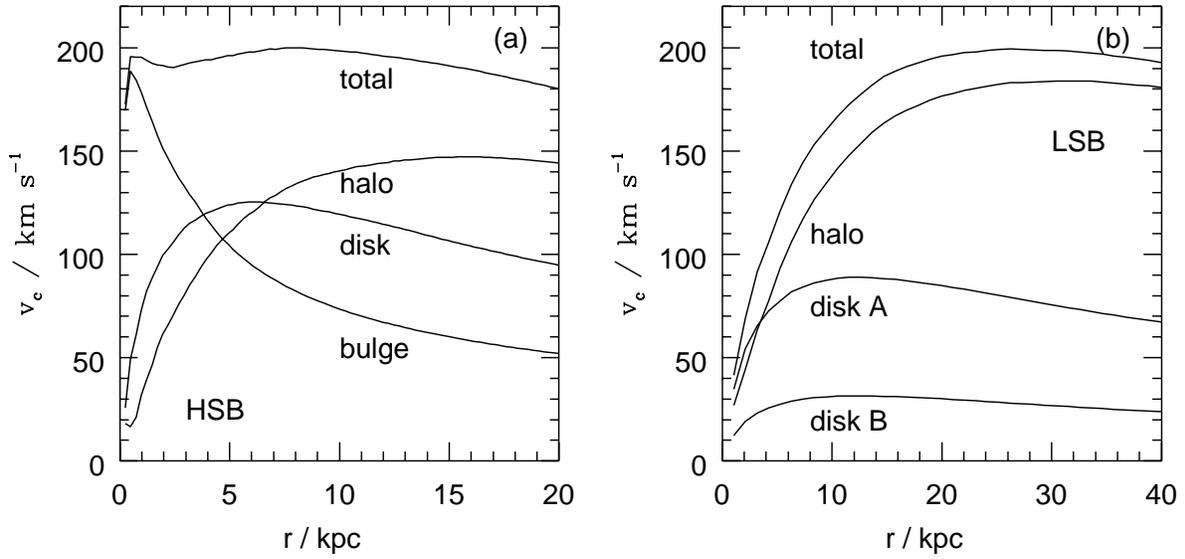}
\caption{
The curves show the contributions from stars and dark matter to the
total rotational velocity of the disk within (a) the HSB galaxy and
(b) the two LSB galaxies, one with a less massive disk but with the same
peak rotational velocity.
}
\label{f:pro}
\end{figure}

The force softening is $0.1r_d$ for the star particles and $0.3r_h$
for the halo particles. Their disks are stable and they remain in
equilibrium when simulated in isolation. Discreteness in the halo
particles causes the disk scale height to increase with time as
quantified in Section 3 for the HSB galaxy.  This increase is
consistent with analytic calculations from Lacey \& Ostriker (1988).

We illustrate the effect of a single impulsive encounter on each of
our model disks in Figures 2, 3 and 4.  At time t=0 we send a
perturbing halo of mass $2\times 10^{12}M_\odot$ perpendicular to the
plane of the disk at an impact parameter of 60 kpc and velocity of
1500 $\kms$.  This encounter would be typical of that occurring in a
rich cluster with a tidally truncated $L_*$ elliptical galaxy near the
cluster core.  Any one galaxy in the cluster will suffer several
encounters stronger than this since the cluster formed.  Although we
simulate a perpendicular orbit here, we do not expect the encounter
geometry to make a significant difference since the difference between
direct and retrograde encounters will be relatively small {\it i.e.}
$V >> v_c$.

At t=0.1 Gyrs after the encounter, the perturber has moved 150 kpc
away, yet a visible disturbance in the disk is hardly
apparent. After 0.2 Gyrs, we can begin to see the response to the
tidal shock as material is torn from the disk into extended tidal
arms.  Even at this epoch there is a clear difference to the response
of the perturbation by the HSB and LSB galaxies.  
After 0.4 Gyrs, the LSB galaxies are
dramatically altered over their entire disks and a substantial fraction
of material has been removed past their tidal radii.  Remarkably, the
central disk of the HSB galaxy remains intact and only the outermost
stars have been strongly perturbed.  A Gyr after the encounter the HSB
disk remains stable whereas the LSB disks are highly distorted.
The model with the more massive disk 
has undergone a strong bar instability.  
The second LSB model with a lighter disk responds in a
similar fashion.  The disk undergoes strong distortions from the
encounter and the same amount of material is tidally removed. However,
the lower mass surface density of this disk has suppressed the bar
instability, confirming the results of Mihos \etal (1997).

\begin{figure}
\centering
\epsfxsize=\hsize\epsffile{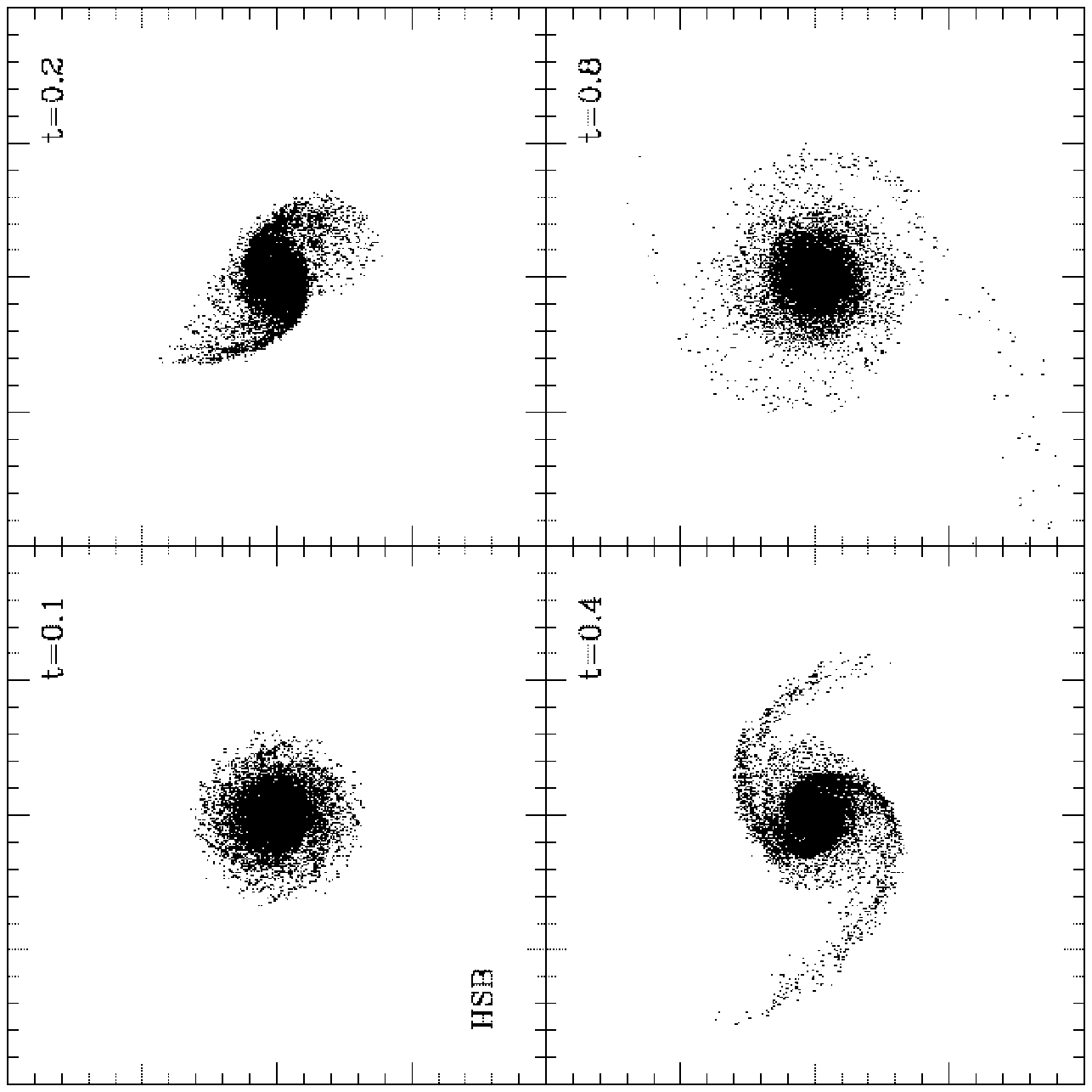}
\caption{
Snapshots of the distribution of disk stars from a HSB galaxy 
after a single high-speed encounter with a massive galaxy.  
Each frame is 120 kpc on a side and encounter takes place
perpendicular to the disk at the box edge (60 kpc).
}
\label{f:points_hsb}
\end{figure}

\begin{figure}
\centering
\epsfxsize=\hsize\epsffile{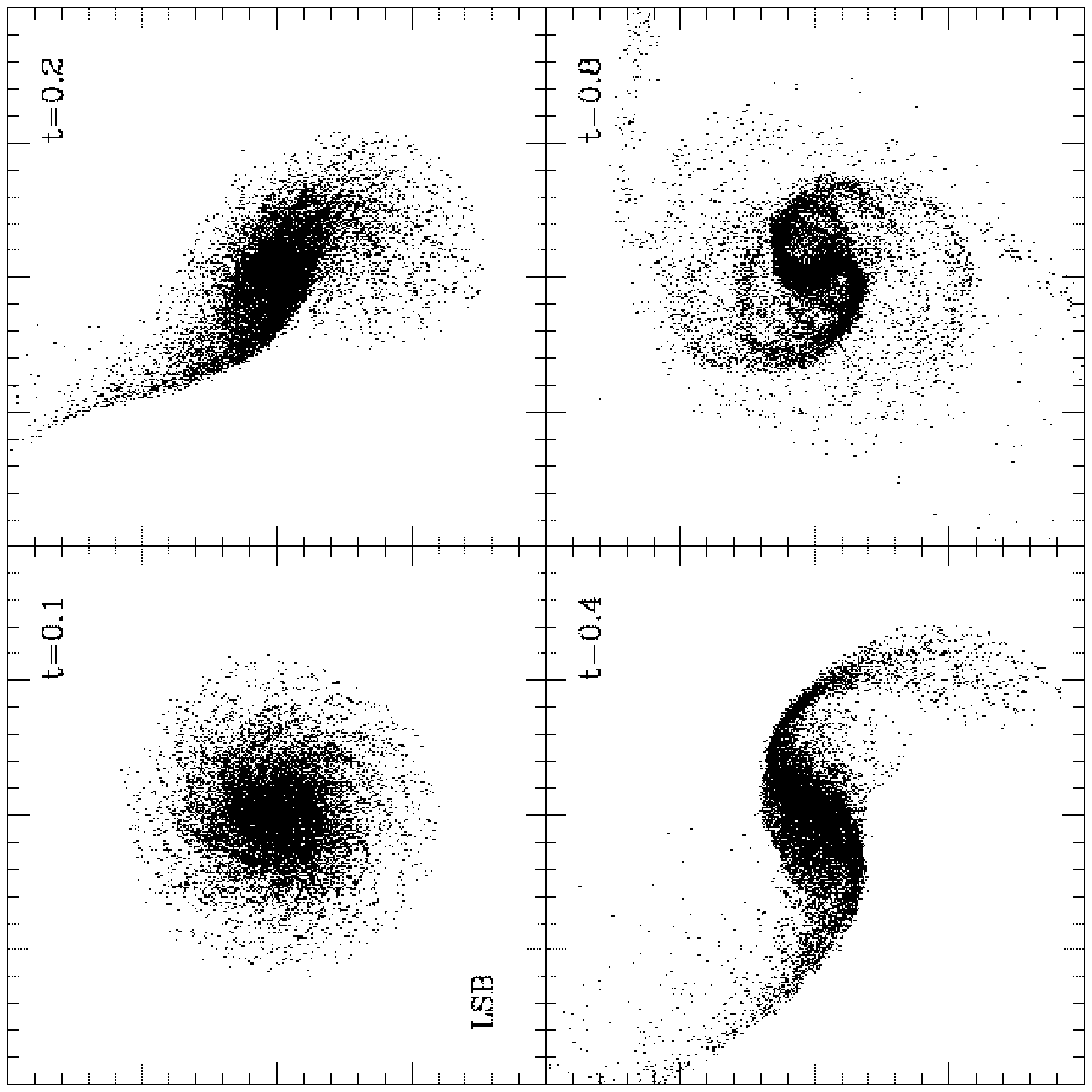}
\caption{
Snapshots of the distribution of disk stars from the LSB galaxy
with a heavy disk (A in Figure 1b)
after a single high-speed encounter with a massive galaxy.  
Each frame is 120 kpc on a side and the encounter takes place
perpendicular to the disk at the box edge (60 kpc).
}
\label{f:points_lsb}
\end{figure}

\begin{figure}
\centering
\epsfxsize=\hsize\epsffile{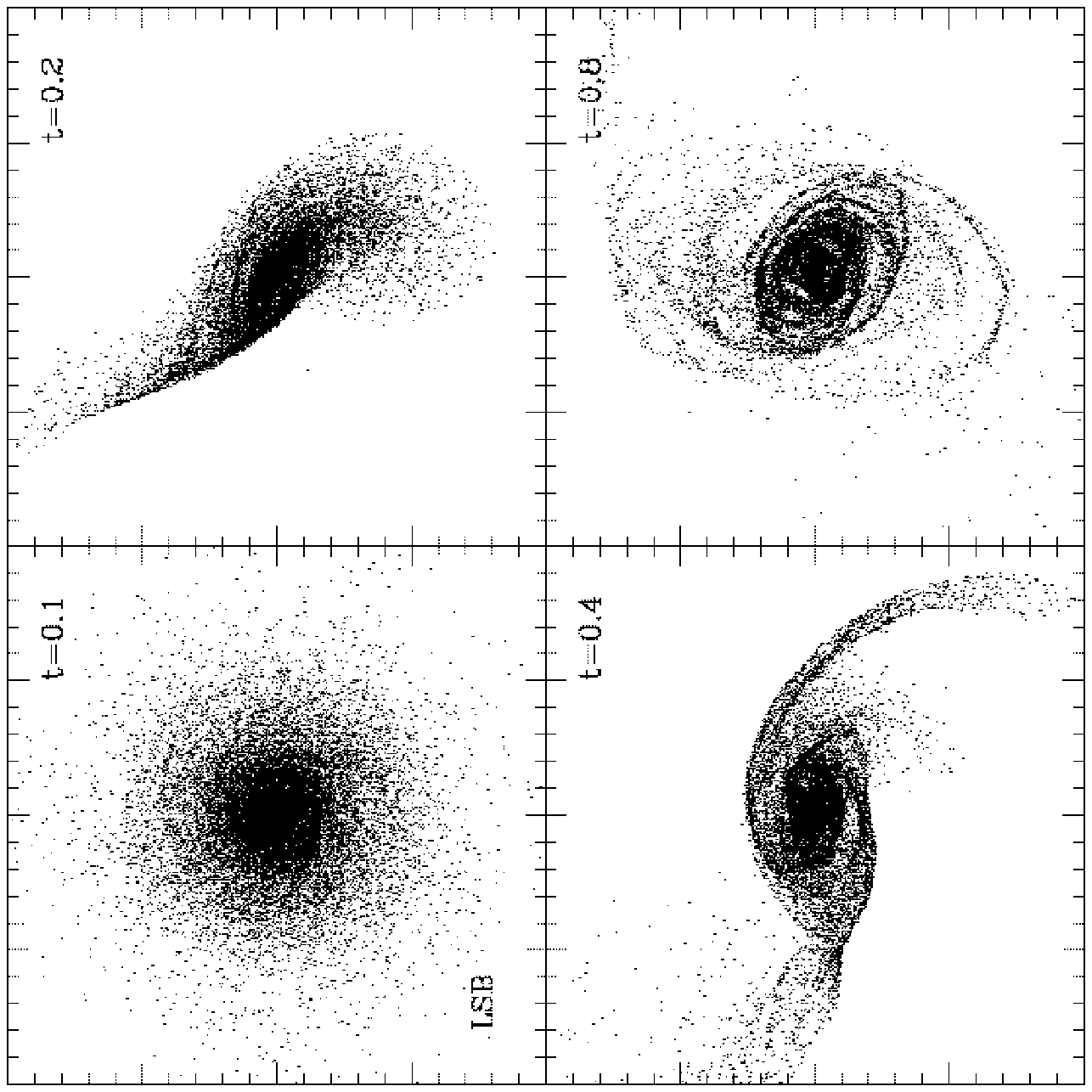}
\caption{
Snapshots of the distribution of disk stars from the LSB galaxy
with a light disk (B in Figure 1b)
after a single high-speed encounter with a massive galaxy.  
Each frame is 120 kpc on a side and the encounter takes place
perpendicular to the disk at the box edge (60 kpc).
}
\label{f:points_lsb}
\end{figure}

\section{Simulating disk evolution within a hierarchical universe}

Previous simulations of tidal shocks and galaxy harassment focussed
upon the evolution of disk galaxies in static clusters with
substructure represented by softened potentials with masses drawn from
a Schechter function (Moore \etal 1996a \& 1998). We shall 
use a more realistic approach of treating the perturbations by
following the growth of a cluster within a hierarchical cosmological
model. The cluster was extracted from a large CDM simulation of a
critical universe within a $50$ Mpc box and was chosen to be virialised
by the present epoch.  (We assume a Hubble constant of 100 \kmsmpc \
.)  Within the turn-around region there are $\sim 10^5$ CDM particles
of mass $10^{10}M_\odot$ and their softening length is 10 kpc.  At a
redshift $z=0$ the cluster has a one dimensional velocity dispersion
of $700\kms$ and a virial radius of 2 Mpc. The tidal field from the
mass distribution beyond the cluster's turn-around radius is simulated
with massive particles to speed the computation. (The cluster used
here is a low resolution version of the cluster analysed in Ghigna \etal.)

Our aim is to select dark matter halos that are likely to host spiral
galaxies and that enter the cluster as it is forming. These halos will
be replaced with the high resolution, stable models and the simulation
continuued to the present epoch. A similar technique was used recently
by Dubinski (1998) in order to study the formation of central cluster
galaxies.  We use the parallel treecode PKDGRAV (Stadel \etal 1998)
that has periodic boundary conditions, accurate force resolution and a
multi-stepping algorithm.  This is vital in this simulation since we
must obtain the correct dynamics on sub-kpc scales in a galaxy that is
feeling the gravitational tidal field from regions several megaparsecs
away.

There are several possible sources of artificial heating that arise
from finite timestepping and artificially large dark matter particle
masses.  Other numerical problems include artificial disk heating by
the general background of particles and the dissolution of small scale
substructure by the tidal shocks, low resolution and force softening.

\subsection{Timestepping, resolution and particle discreteness}

In order to model the dynamics of star particles within the disk and
the growth of the cluster at the same time an efficient multi-stepping
algorithm is needed. Most of the particles in the cosmological volume
have larger softening and lower velocities than the high resolution
galaxy, therefore a fixed timestep would be inefficient.  The
softening length for star particles in the disk of the model HSB
galaxy is $\sim 200$ pc, whilst the CDM particles in the main cluster
have 10 kpc softening.  At relative velocities of several thousand
$\kms$, some particles require timesteps of order $\sim 10^5$ years
and thus require more than 50,000 steps in total.  The multistepping
criteria of PKDGRAV is based on the local acceleration, therefore it
is better suited for following encounters than is a velocity
criteria. Furthermore, a velocity criteria is inefficient for circular
orbits within isothermal potentials since all the disk particles would
be on the same timestep regardless of their local density.  To test
the multistepping and accuracy of the force calculation, we placed the
model galaxy in a void, 10 Mpc away from the forming cluster at
z=0.5. The galaxy evolved in relative isolation for half a Hubble
time, yet remained very stable, although the disk scale height
increases slowly due to discreteness as quantified in Figure 4. Even
with 50,000 halo particles within 10 disk scale lengths, the disk
vertical scale height more than doubled over the 5 Gyr integration.

\begin{figure}
\centering
\epsfxsize=\hsize\epsffile{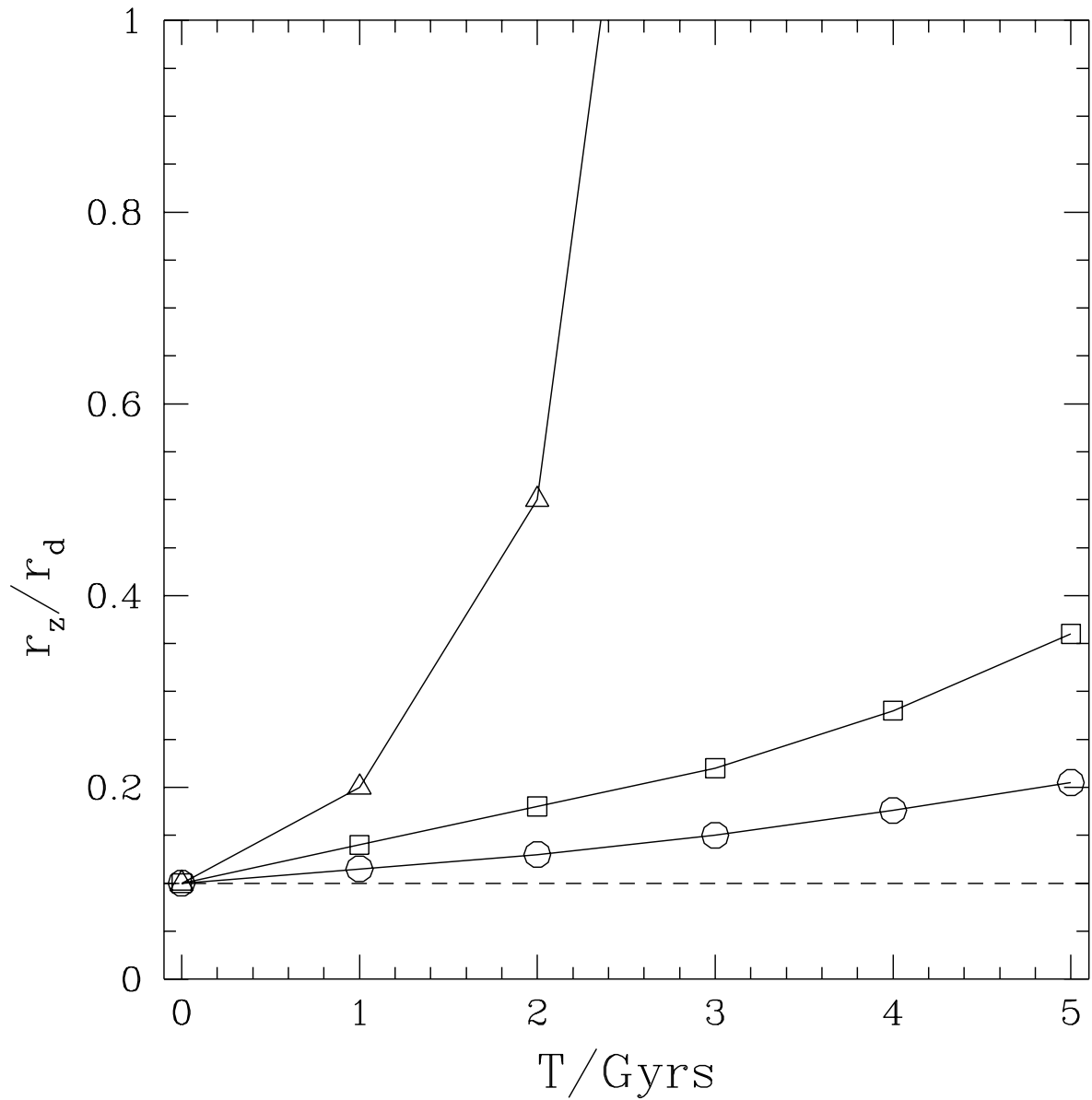}
\caption{
The vertical scale height, $r_z$, of the disk in units of the initial 
disk scale length, $r_d$, measure at $r_d$ plotted against time. 
The circles are data for the HSB galaxy placed in a void. The squares 
and triangles show one of the HSB and LSB galaxies respectively, that 
enters the cluster at z=0.5.
}
\label{f:disk}
\end{figure}

A long standing problem in ``low-resolution'' dissipationless N-body
simulations is the dissolution of substructure. When large
cosmological volumes are simulated, virialised halos with less than
$\lsim 10^5$ particles contain very little substructure.  This is due
to tidal shocks disrupting the heavily softened halos that fall into larger
systems (Moore \etal 1996b).  The absence of halos-within-halos will
artificially reduce the effects of harassment in our simulation since
the full mass spectrum of perturbing clumps will not be present.
However, harassment is dominated by the effects of several strong
encounters with large halos of mass $\gsim L*\equiv v_c=220\kms$.  All
halos more massive than this are resolved by our simulation outside of
the cluster, and some fraction will survive for at least a crossing
time within the cluster.

At the final time within the cluster's virial radius there are just 6
surviving substructure halos with circular velocities larger than
$220\kms$. (N.B. When this cluster was simulated with $\sim 10^6$ dark
matter particles and 5 kpc softening, we found 17 halos with $v_c
\gsim 220\kms$ within the virial radius.)  We find that the chaos of
cluster formation is the time when most damage is caused to the disk
galaxies. At a redshift $z\sim 0.5$ many smaller halos are streaming
together along filaments at high velocities - encounters with these
halos wreak havoc with the disks of LSB galaxies, and including the
subsequent encounters with galaxies within the cluster will only add
to the disk heating.  In agreement with the analysis of Ghigna \etal,
none of the galaxies suffer a single merger whilst orbiting within the
cluster.

\subsection{Results}

Between a redshift z=2 to z=0.5 we follow the merger histories of
several candidate dark matter halos from the cosmological simulation
that end up within the cluster at later times.  We select three halos
with circular velocities $\sim 200\kms$ that have suffered very little
merging over this period and would therefore be most likely to host
disk galaxies. We extract these halos from the simulation at z=0.5 and
replace the entire halo with the pre-built high resolution model
galaxies.  We rescale the disk and halo scale lengths by $(1+z)^{-1}$
according to the prescription of Mao \etal (1998) to represent the
galaxies entering the cluster at higher redshifts.  This theoretical
prescription is based on modeling disk formation within a hierarchical
universe. Although observational evidence for this behaviour is lacking
(Lilly \etal 1998), it will only serve to make disks at
higher redshifts more stable to harassment.

On a 64 node parallel computer, each run takes several hours; three
runs were performed in which the halos were replaced with the HSB disks
and a further six runs using the LSB disks from Figure 1. 
Figure 6 shows the results of one of the LSB simulations from a redshift
z=0.5 to the present day \-- this evolution is typical of all six runs.
The colours show the local density of CDM particles on a
scale of $10-10^6\rho_{crit}$ and the size of each image is a comoving
10 Mpc.  At z=0.5, the cluster is only just starting to form from a
series of mergers of several individual group and galaxy sized halos -
the small halo that we have replaced with an LSB galaxy at z=0.5 is
highlighted in the first two images.  The green points show the stars
within the stellar disk that are barely visible on this scale. The
cluster quickly virialises, although several dark matter clumps
survive the collapse and remain intact orbiting within the clusters
virial radius.  Between $z=0.4\--0.3$ the model galaxy receives a
series of large ``tidal shocks'' from the halos that are assembling
the cluster.

Once the galaxy enters the virialised cluster, it continues to suffer
encounters with infalling and orbiting substructure.  By a redshift
z=0.1, most of the stars have been stripped from the disk and now
orbit through the cluster - closely following the rosette orbit of the
parent galaxy. The final orbit of this run has apocenter of 1000 kpc
and pericenter of 150 kpc.  Of the LSB galaxy runs, between 50\% and
90\% of the stars were harassed from the disk, whereas the stellar
mass loss in the HSB runs was between 1\% and 10\%.  We find no
discernable difference between the stellar mass loss or the
kinematical remnants of the two different LSB models in Figure 1(b).
The tidal encounters are so strong and frequent that the additional 
stability provided by a dark matter dominated disk is not apparent.

\begin{figure}
\centering
\epsfxsize=\hsize\epsffile{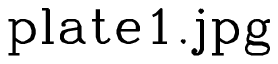}
\caption{
Snapshots of the particle distribution
within a comoving 10 Mpc box centered on the forming cluster. The
redshifts of each frame are indicated.  The colors show the local
density of dark matter on a scale of $1\--10^6$ times the mean
density.  The green particles are the star particles from the disk of
the high resolution model galaxy. Initially the stars are confined to
the disk, but at z=0 they are spread along the orbital path of
the galaxy which has apocenter of 1000 kpc and pericenter of 250kpc.
The clusters virial radius at z=0 is $\sim 2$ Mpc.
}
\label{f:final3}
\end{figure}


\subsection{The final stellar states}

Only three different orbital realisations of each LSB model were
carried out, therefore we cannot comment on correlations between
properties, but some general remarks about the kinematics of the
stellar remnants can be made.  The final stellar systems are prolate,
with shapes supported by velocity anisotropy, similar to the remnants
of harassed 
Sc-Sd galaxies analysed in Moore \etal (1998).  Figure 7(a) shows
the initial and final surface density of stars from the LSB
galaxy. Within $\sim$ 5 kpc, the remnants are well fitted by
exponentials with scale lengths in the range $\sim 1.5-2.5$ kpc, a
significant decrease from their initial values.  
Even though 50-90\% of the stars have been tidally stripped, the central
surface brightness increases by up to 2 magnitudes.  This increase in
the central stellar density results from an increase in the random motions of
stars by the heat input from harassment - phase space density is
conserved. The dark matter particles are initially on more radial 
orbits than the disk stars, therefore they can be stripped from deeper within
the potential halo if they are caught at apocenter.
This results in a larger fraction of dark matter being removed,
even from within the optical extent of the galaxies and the final
stellar mass to dark matter ratios decrease from 16 to $\sim 2-5$.

\begin{figure}
\centering
\epsfxsize=\hsize\epsffile{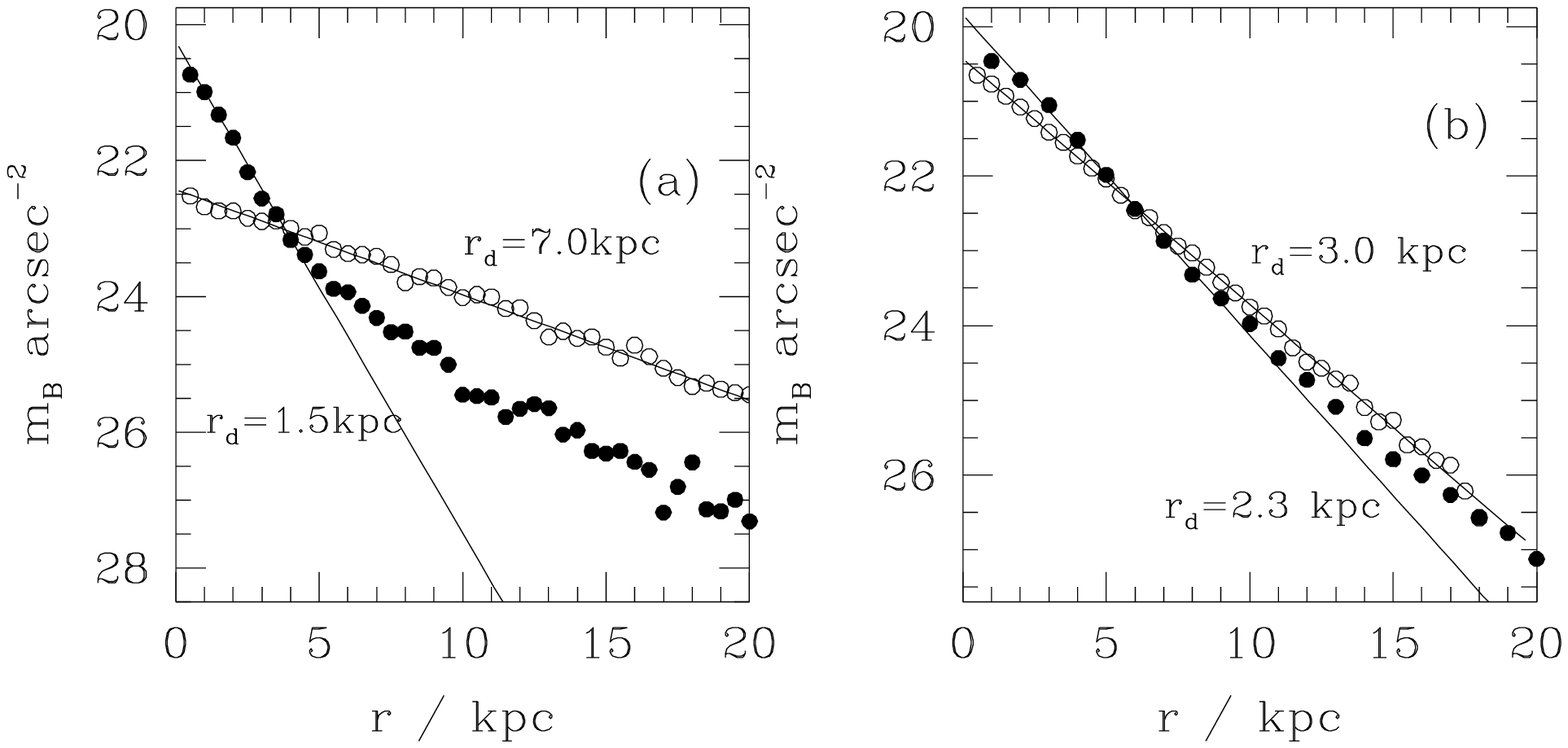}
\caption{The open circles show the initial surface density of disk
stars in (a) the LSB model galaxy and (b) the HSB model galaxy.  The open
circles show the surface density at the final time after evolving within
the cluster. The solid lines show exponential disk fits with 
the indicated scale lengths.
}
\label{f:surface}
\end{figure}

Tidal stripping combined with the additional fading from several Gyrs of
stellar evolution will leave the remnants faint and diffuse.
Combining these results with our simulations of fainter Sc-Sd galaxies in
clusters, we expect that the luminosity function of dwarf spheroidals
(dSph - frequently called dE's) in clusters should reflect the 
luminosity function of bulgeless
Sc-Sd spirals and LSB galaxies that have entered the cluster.
Therefore, if a substantial population of
luminous LSB galaxies exist, we should find their post-genitors in
clusters as a large population of diffuse low luminosity spheroidals with
low mass to light ratios.

The evolution of the HSB models contrasts sharply with the LSB models.
These galaxies lose only a small fraction of their stars and
they remain in a disk configuration.  The scale height of their disks
increases by a factor of 2-4 over that sustained by discreteness
effects.  The increase in the stellar velocity dispersion also leads
to a small increase in the central surface brightness by about half a
magnitude (see Figure 7(b)).  Within about 10 kpc the disks are well fitted by
exponentials with scale lengths that are slightly smaller by $\sim
20$\% than their initial values.  Beyond this region we find an excess
in surface brightness over a single exponential disk fit; at 20 kpc
this is over one magnitude.  The final disks do not have any obvious
spiral features and if the gas is removed via ram-pressure stripping,
we speculate that these galaxies would be identified as S0 galaxies in
present day clusters.

\section{What determines the stability or instability of disks in clusters?}

The parameters for the ``typical'' HSB and LSB galaxies that we
constructed are different in many aspects. It is not clear which
parameter, or combination of them maintains the stability of the HSB
galaxy. We performed a series of tests to isolate the effects of
the three important structural parameters; the surface mass density of
the disk, the disk scale length and the depth of the potential well
provided by a dark halo.  (A prominent stellar bulge will have a
similar effect as a dark matter halo with a very small core radius.)

We construct 7 model galaxies that have equilibrium disk + halo
components. The disks have scale lengths of either 3 kpc or 7 kpc and
a dark halo with core radius 1 kpc or 10 kpc which provides a peak
rotational velocity of $200 \kms$. All possible combinations of these
parameters yield 4 models and we construct a further 3 models in which
we vary only the disk mass. The model parameters are summarised in Table 1.
The contribution to the rotation curve $v_c=\sqrt{GM/r}$ of the halo
and stellar components of each model are plotted in Figure 8.
The entire simulation is re-run seven times with each model on the
same orbit, thus they suffer identical harassment histories within the
cluster.  The dashed and dotted curves in Figure 8 show the final
rotation curves of the stellar and halo components of all seven models.

\begin{table}
\begin{center}
\begin{tabular}{cccccc}
\hline
Model & $r_d$ & $r_h$ & $M_{halo}/M_{disk}<30$ kpc& stars & dark matter    \\ \hline

A   & 7 kpc         & 10 kpc     &   16      & 57\%       & 92\%      \\
B   & 7 kpc         & 1 kpc      &   16      & 23\%       & 40\%      \\
C   & 7 kpc         & 1 kpc      &    7      & 20\%       & 35\%      \\
D   & 7 kpc         & 1 kpc      &   57      & 22\%       & 41\%      \\ 
E    & 3 kpc        & 10 kpc      &   16      & 1\%       & 93\%      \\
F    & 3 kpc        & 1 kpc       &   16      & 1\%      & 35\%      \\  
G    & 3 kpc        & 10 kpc      &   57      & 2\%       & 60\%      \\ \hline 


\end{tabular}
\caption{Models to test disk stability as a function of potential depth and 
disk structure. Columns 2 and 3 are the disk scale lengths and halo core radii,
column 4 gives the dark to stellar mass ratio within 30 kpc and
columns 5 and 6 show the percentage of stars and dark matter
that are stripped from within a radius of 30 kpc by the final time.}
\label{t:one}
\end{center}
\end{table}

\begin{figure}
\centering
\epsfxsize=\hsize\epsffile{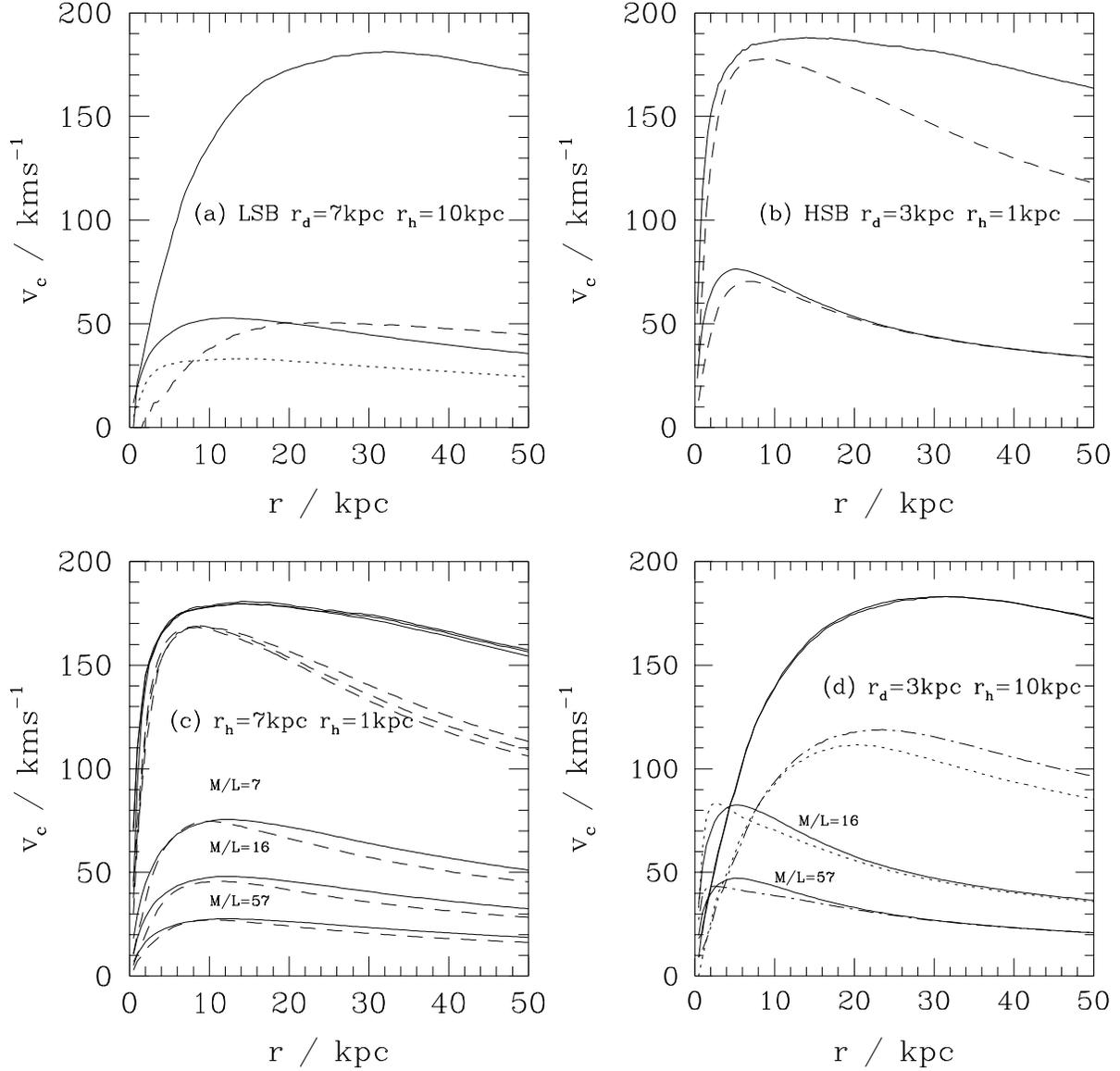}
\caption{The contribution of the stars and dark matter to the
rotational velocity of the seven test galaxies, all plotted in
physical coordinates. The initial mass distributions are shown
by the solid curves whilst the final mass distributions are shown as
dotted or dashed curves.  The long dashed and dotted curves in panel
(a) show the halo and disk distributions respectively at the
final time.  The other galaxies in panels (b)-(d) lose much less mass
and it should be relatively clear which component the broken
curves respresent at the final time.
}
\label{f:4}
\end{figure}

The standard lore is that LSB disks inhabit low density halos (rising
rotation curves) and HSB disks live in concentrated halos (flat
rotation curves).  These models would correspond to A and F
respectively in Table 1; their final mass distributions are plotted in
Figure 8(a) and 8(b). The behaviour of these models is similar to the
two ``standard'' models discussed in the previous section.

We know that a disk is highly unstable to mass loss, disruption and
morphological transformation if both the halo core radius and disk
scale lengths are large. Which parameter is most important?  The
answer is that they both are.  From Table 1 we can see that a galaxy
with a scale length of 3 kpc is stable against mass loss independent
of the halo core radii, although their final disk 
scale heights are significantly
higher in the case of an extended halo.  Furthermore, an extended disk
that was highly unstable in a soft potential can be stabilised in a
concentrated potential, although they suffer significant mass loss
and large amounts of internal heating.

The core radius of the mass distribution makes a large difference in
the extent of the final mass distribution.  Models with $r_h=10$ kpc
and $r_h=1$ kpc (with the small disk), had final tidally limited radii
of $\sim$ 30 kpc and 50 kpc.

Models B C and D are identical except for their disk mass which vary
by a factor of eight. The stellar and dark matter mass loss is very
similar between all these models and we cannot find any significant
distinguishing feature between the final stellar systems. Although
the disks lose over a fifth of their mass, the final objects are
rotationally supported but suffer a lot of heating. Within a disk
scale length, the initial and final surface density profiles are
similar to the initial conditions, although model C ends up slightly
more concentrated than model D.  The deeper potentials have stopped
the transformation in dSph galaxies but the remaining disks have large
scale heights ($\gsim 2$ kpc) and they all look like extremely puffed
up low surface brightness S0 galaxies.

\section{Summary and discussion}

We have identified halos within a cosmological simulation that have
quiet merger histories and that enter a cluster environment at a redshift
$z= 0.5$.  At this epoch we replace the halos with high resolution
model spiral galaxies with either LSB or HSB disks and continue the
simulation to the present day.  This technique is a simple and
powerful tool for studying the morphological evolution of galaxies in
different environments.  


The response of a disk galaxy to tidal shocks is governed primarily by
the concentration of the mass distribution that encompasses it and the
disk scale length.  LSB galaxies have slowly rising rotation curves,
large disk scale lengths and they cannot survive the chaos of cluster
formation; gravitational tidal shocks from the merging substructure
literally tear these systems apart, leaving their stars orbiting
freely within the cluster and providing the origin of the
intra-cluster light.

Recent observations of individual planetary nebulae within clusters,
but outside of galaxies, lends support to this scenario.  Estimates of
the total diffuse light within clusters, using CCD photometry
(Bernstein \etal 1995, Tyson \& Fischer 1995) or the statistics of
intra-cluster stars (Theuns \& Warren 1997, Feldmeier \etal 1998,
Mendez \etal 1998, Ferguson \etal 1998), ranges from 10\% to 45\% of
the light attached to galaxies. Presumably, these stars must have
originated within galactic systems. The integrated light within LSB
galaxies may be equivalent to the light within ``normal'' spirals
(Bothun, Impey \& McGaugh 1997, and references within).  This is
consistent with the entire diffuse light in clusters originating from
harassed LSB galaxies. 

Models for the formation of LSB galaxies typically assume that they
form within dark matter halos with high spin parameters (Dalcanton
\etal 1997, Jiminez \etal 1998).  Recent work by Lemson \& Kauffmann
(1998) demonstrates that the properties of dark matter halos,
including spin parameter, are independent of environment.  Therefore,
the initial distribution of LSB galaxies should be unbiased with
respect to the local overdensity, and the quantity of intra-cluster
light within clusters may provide an upper limit to the maximum 
amount of light that can exist in LSB galaxies in the universe.

High surface brightness disk galaxies and galaxies with luminous
bulges have steep mass profiles that give rise to flat rotation curves
over their visible extent.  The orbital time within a couple of disk
scale lengths is short enough for the disk to respond adiabatically to
rapid encounters.  Tidal shocks cannot remove a large amount of
material from these galaxies, nor transform them between morphological
types, but will heat the disks and drive instabilities that can funnel
gas into the central regions (Lake \etal 1998, Gnedin 1999). A few Gyrs after
entering a cluster, their disks are thickened and no spiral features
remain.  If ram-pressure is efficient at removing gas from disks, we
speculate that these galaxies will evolve into S0's.  Since
the harassment process and ram-pressure stripping are both more
effective near the cluster centers, we expect that a combination of
these effects may drive the morphology--density relation within
clusters.

\acknowledgments

We thank Stacy McGaugh for many useful comments and suggestions on
this work. Computations were carried out using COSMOS, the HEFCE and
PPARC funded Origin 2000 as part of the Virgo Consortium. Ben Moore is
a Royal Society research fellow.

\baselineskip=8pt

\clearpage
\vskip 1.0truein

\noindent{\bf References}


\pp Abadi M., Moore B. \& Bower R.G. 1998, im preparation.

\pp Bernstein, G.M., Nichol R.C., Tyson J.A., Ulmer M.P. \& Wittman D.
1995, {\it A.J.}, {\bf 110}, 1507.

\pp Binggeli, B., Sandage, A. and Tammann G.A.  1988, {\it
Ann. Rev. Astr. Ap.}, {\bf 26}, 509.

\pp Binggeli B., Tammann, G.A. and Sandage, A. 1987, {\it A.J.}, {\bf
94}, 251.

\pp de Blok W.J.G. \& McGaugh S.S. 1996, {\it Ap.J.Lett.}, {\bf 469}, L89.

\pp de Blok W.J.G. \& McGaugh S.S. 1997, {\it M.N.R.A.S.}, {\bf 290}, 533.

\pp Bothun G.D., Schombert J.M., Impey C.D., Sprayberry D. \& McGaugh S.S.
1993, {\it A.J.}, {\bf 106}, 530.

\pp Bothun G.D., Impey C. \& McGaugh S. 1997, {\it P.A.S.P.}, {\bf 109}, 745.

\pp Butcher, H. and Oemler, A. 1978, {\it Ap.J.}, {\bf 219}, 18.

\pp Butcher, H. and Oemler, A. 1984, {\it Ap.J.}, {\bf 285}, 426.

\pp Byrd G. and Valtonen M. 1990, {\it Ap.J.}, {\bf 350}, 89.


\pp Couch W.J., Barger A.J., Smail I., Ellis R.S. \& Sharples R.M. 1998,
{\it Ap.J.}, {\bf 497}, 188. 

\pp Dalcanton J.J., Spergel D.N. \& Summers FJ. 1997. {\it Ap.J.}, 
{\bf 482}, 659.

\pp Dressler, A, Oemler, A., Butcher, H. and Gunn, J.E.  1994a, {\it
Ap.J.}, {\bf 430}, 107.

\pp Dressler A., Oemler A., Couch W.J., Smail I., Ellis R.S., Barger A., 
Butcher H., Poggianti B.M., Sharples R.M. 1998, {\it Ap.J.}, {\bf 490}, 577.

\pp Dubinski J., 1998, {\it Ap.J.}, {\bf 502}, 141.

\pp Feldmeier J, Ciardullo R. \& Jacoby G. 1998, {\it Ap.J.}, {\bf 503}, 109.

\pp Ferguson H.C., Tanvir N.R. \& von Hippel T. 1998, {\it Nature}, {\bf 391} 461.

\pp Ghigna, S., Moore, B., Governato, F., Lake, G., Quinn, T. \& Stadel, J.
1998, {\it M.N.R.A.S.}, {\bf 300}, 146.

\pp Gnedin, O. 1999, {\it Ap.J.}, submitted.
 
\pp Gunn J.E. \& Gott J.R. 1972, {\it Ap.J.}, {\bf 176}, 1.

\pp Hernquist, L. 1993, {\it Ap.J.Suppl.}, {\bf 86}, 389.

\pp Icke, V. 1985, {\it Astr. Ap.}, {\bf 144}, 115-23.

\pp Jimenez R., Padoan, P., Matteucci F. \& Heavens A.F., 1998, {\it MNRAS}, {\bf 299}, 515.

\pp Lacey C. \& Ostriker J.P. 1985, {\it Ap.J.}, {\bf 299}, 633.

\pp Lake, G., Katz, N. and Moore, B. 1998, {\it Ap.J.}, {\bf 495}, 152.

\pp Lavery R.J. \& Henry J.P. 1988, {\it Ap.J.}, {\bf 330}, 596.

\pp Lavery R.J. \& Henry J.P. 1994, {\it Ap.J.}, {\bf 426}, 524.

\pp Lemson G. \& Kauffmann G. 1997, {\it MNRAS}, submitted, astro-ph/9710125.

\pp Lilly S. \etal 1998 {\it Ap.J.}, {\bf 500}, 75.

\pp Mao S., Mo. H.J \& White S.D.M. 1998, {\it M.N.R.A.S.}, {\bf 297}, 71. 

\pp Mendez R.H., Guerrero M.A., Freeman K.C., Arnaboldi M., Kudritzki R.P., 
Hopp U., Capacciolo M. \& Ford H. 1997, {\it Ap.J.Lett.}, {\bf 491}, 23.

\pp Mihos J.C., McGaugh S.S. \& de Blok W.J.G. 1997, {\it Ap.J.Lett.}, 
{\bf 477}, L79.

\pp Mo H.J., McGaugh S.S. \& Bothun G.D. 1994, {\it M.N.R.A.S.}, 
{\bf 267}, 129.

\pp Moore, B., Katz, N. and Lake, G. 1996b, {\it Ap.J.}, {\bf 457},
455.

\pp Moore, B., Katz N., Lake G., Dressler, A. and Oemler, A. 1996a, {\it
Nature}, {\bf 379}, 613.

\pp Moore, B., Lake, G. \& Katz, N. 1998, {\it Ap.J.}, {\bf 495}, 139.

\pp Persic M. \& Salucci P., 1997, {\it Dark and visible matter 
in galaxies}, {\bf ASP Conference series, 117}, ed. M. Persic
P. Salucci.

\pp Pickering T.E., Impey C.D., Van Gorkom J.H. \& Bothun G.D. 1997,
{\it A.J.}, {\bf 114}, 1858.

\pp Stadel J., Quinn T. \& Lake G. 1998, in preparation.

\pp Theuns T. \& Warren S.J. 1997, {\it M.N.R.A.S.}, {\bf 284}, L11.

\pp Thompson, L.A. and Gregory, S.A. 1993, {\it A.J.}, {\bf 106},
2197.

\pp Tyson J.A. \& Fischer P. 1995, {\it Ap.J.Lett.}, {\bf 446}, L55. 

\pp Valluri, M. and Jog, C. J. 1991, {\it Ap.J.}, {\bf 374}, 103.

\pp Zwaan M.A., van der Hulst J.M., de Blok W.J.G. \& McGaugh S.S. 1995, {\it M.N.R.A.S.}, 
{\bf 273}, L35.

\baselineskip=14pt




\baselineskip=14pt

\bigskip

\end{document}